# Do interdisciplinary research teams deliver higher gains to science?[1]

*Giovanni Abramo[a], Ciriaco Andrea D'Angelo[a,b], Flavia Di Costa[a]*

[a] Laboratory for Studies in Research Evaluation
Institute for System Analysis and Computer Science (IASI-CNR)
National Research Council of Italy

[b] Department of Engineering and Management
University of Rome "Tor Vergata"

**Abstract**

The present paper takes its place in the stream of studies that analyze the effect of interdisciplinarity on the impact of research output. Unlike previous studies, in this study the interdisciplinarity of the publications is not inferred through their citing or cited references, but rather by identifying the authors' designated fields of research. For this we draw on the scientific classification of Italian academics, and their publications as indexed in the WoS over a five-year period (2004-2008). We divide the publications in three subsets on the basis the nature of co-authorship: those papers coauthored with academics from different fields, which show high intensity of inter-field collaboration ("specific" collaboration, occurring in 110 pairings of fields); those papers coauthored with academics who are simply from different "non-specific" fields; and finally co-authorships within a single field. We then compare the citations of the papers and the impact factor of the publishing journals between the three subsets. The results show significant differences, generally in favor of the interdisciplinary authorships, in only one third (or slightly more) of the cases. The analysis provides the value of the median differences for each pair of publication subsets.

---

# 1. Introduction

The possibilities of scientific gains through interdisciplinary research (IDR) are of increasing interest to both academics and policy-makers. In 2015, the journal Nature dedicated a special issue[2] to analyzing and debating "how scientists and social scientists are coming together to solve the grand challenges of energy, food, water, climate and health". Earlier, in 2011, Wagner et al. carried out a full review of studies on interdisciplinarity, examining the different approaches to understanding and measuring IDR. Their study provided "a more holistic view of measuring IDR, although research and development is needed before metrics can adequately reflect the actual phenomenon of IDR". Wagner et al. found that among the different quantitative measures of IDR, the ones most frequently studied and used were the bibliometric measures (co-authorships, co-inventors, collaborations, references, citations and co-citations). The same authors also criticized the persistent gap in understanding the social dynamics for integrating knowledge from IDR.

The most common method used for measuring the IDR phenomenon is citation analysis of publications. The occurrence of citations to publications belonging to a range of different scientific fields other than those of the publication citing is considered as signal of possible interactions or integration between the different fields. One of the most analyzed aspects is the effect of interdisciplinarity on the impact of the research products. Various studies have investigated this tie (Steele and Stier, 2000; Rinia et al., 2001; Levitt and Thelwall, 2008; Larivière and Gingras, 2010; Yegros-Yegros et al., 2015; Wang et al., 2015; Chen et al., 2015). However the results are often contrasting, in part because of resorting to different indicators for measuring IDR (Wang et al., 2015).

In the following, we summarize the methodological approaches and findings of the most recent studies.

Yegros-Yegros et al. (2015) analyzed the effect of IDR on the citation impact of individual publications in four different scientific fields. First, the authors measured the disciplinary diversity in the references of a publication: variety (the number of WoS subject categories cited), balance (the distribution of references over WoS subject categories), and disparity (the cognitive distance of the references). Subsequently they investigated the separate effects of the different aspects of IDR diversity on citation impact. Using multivariate regression analysis, the authors were able to separately consider and evaluate the effect of all three dimensions of IDR (variety, balance and disparity) on citation impact, after accounting for the effects of a wide range of control variables. The link between IDR and citation impact resulted as being quite complex: very low or very high degrees of IDR are associated with a decrease of citation impact, while some middle degree of IDR, described as "proximal interdisciplinarity", shows higher citation impact.

Wang et al. (2015) again used factor analysis to investigate IDR variety, balance, and disparity. In summary, the results are that long-term (13-year) citations i) increase at an increasing rate with variety; ii) decrease with balance, iii) increase at a decreasing rate with disparity.

Chen et al. (2015) observed all journal articles indexed in the WoS in the year 2000 and the corresponding lists of references. The citations were counted as of December 2013. The authors analyze the levels of interdisciplinarity, calculated using the Simpson's Index[3] of two sets of publications: 1) the top 1% most cited articles; 2) the articles in other citation rank classes. The results show that in more than 90% of scientific disciplines (by NSF classification) the publications of the most-cited set have higher levels of interdisciplinarity. The authors thus conclude that IDR is one of the factors capable of producing higher impact knowledge.

---

[2] http://www.nature.com/news/interdisciplinarity-1.18295, last accessed on 9 November 2016.

An alternative approach to studying IDR is to base the citation analysis on the specializations of the co-authors. Such analyses consider that "An interdisciplinary group consists of persons trained in different fields of knowledge (disciplines) with different concepts, terms, methods and data organized by a common effort working on a common problem with continuous intercommunication" (OECD 1972, p. 25-26). However the application of this approach presents serious operational problems (Porter, Cohen et al., 2007). In fact, while it is possible to examine the curricula vitae of the authors, the identification of their fields of research is then exceptionally demanding in terms of time, and requires expert judgment. There are thus very few cases where the method has been applied, such as Schummer (2004) and Porter, Roessner, and Heberger (2008). Indeed these studies are based on small samples of scientists, and the authors themselves note the tedious nature of collecting and processing the data.

The present work once again studies IDR through the identification of co-author specialization. However, to do this, it takes advantage of an unusual feature in the organization of the Italian research system, permitting a massive database with respect to the previous studies. The "advantage" of the Italian system is that under legislative regulation, all university professors must be classified in one and only one scientific field. There are 370 such fields, named Scientific Disciplinary Sectors (SDSs), grouped into 14 University Disciplinary Areas (UDAs). Beginning with this framework, the current authors were then able to develop an algorithm for the disambiguation of the authorships of articles indexed in the WoS (D'Angelo et al., 2011). We are thus able to automatically attribute each publication to its academic authors, for whom the SDSs are already clearly identified. We are therefore able to answer the research question of whether IDR teams produce higher impact outputs (Q1), overcoming the limits of the "co-author specialization approach" encountered by the previous scholars. We can also attempt to answer the further question of whether such teams succeed at publishing their research outputs in more prestigious journals (Q2).

In the next section of the paper we present the methodology of the study and a description of the dataset. Section 3 provides the results of the analysis. In the final section we offer the conclusions and discuss the limits and potential future developments from the study.

## 2. Methods and Data

Our analysis of the relation between IDR teams and the impact of research output is limited to the hard sciences. The bibliometric approach cannot guarantee robust and reliable analyses in the social sciences or humanities, due to the scarce coverage of these areas in the bibliometric databases. Our potential field of observation is thus composed of all Italian professors in the sciences, and of their output indexed in the WoS over the period 2004-2008. We measure the impact of the publications by means of the citations counted on 30 December 2015, and the prestige of the publishing journal by its impact factor (IF) at time of publication. Any publication authored by two or more professors belonging to different SDSs could be the object of analysis, as having been produced by an IDR team. We split IDR publications into two subsets. One subset is made of publications authored by scientists belonging to SDSs that tend to show very high rates of collaboration between their disciplines. We consider the publications resulting from collaboration between such SDSs as the result of "specific" interdisciplinary research. We distinguish these from the publications resulting from what we call "generic" interdisciplinary research, meaning from teams involving disciplines where collaborations are less frequent (second subset).

---

[3] Simpson's Index of Diversity, originally developed in biology, is defined in bibliometrics as $1 - \sum p_i^2$, where $p_i = {x_i}/{X}$; $X = \sum x_i$, and $x_i$ is the number of references to the i-th subject category.

We have already identified such "specific" SDS pairs in a previous work (Abramo et al., 2012), proceeding as follows.

As described, the organization of Italian university personnel provides that each scientist must belong to a specific SDS. Each SDS in turn belongs to a UDA: The sciences consist of 9 UDAs (Mathematics and computer sciences, Physics, Chemistry, Earth sciences, Biology, Medicine, Agricultural and veterinary sciences, Civil engineering, Industrial and information engineering) and 205 SDSs.[4] Using the disambiguation algorithm noted above, the true authors were identified for all publications. Given that each author is associated to an SDS, it is then relatively simple to identify the number of different SDSs represented in the byline, for each indexed publication. We can then carry out the count of the combinations of SDSs that occur with greater frequency. As reported in Annex 2, we identify 110 SDS pairs whose "specific degree of interdisciplinarity"[5] is above 10%. The field of observation is constituted of all the publications authored by professors belonging to the first SDS, for all the pairs in the list. Taking the publications authored by professors of the first SDS of the pair, we then divide their publications in three subsets:

- Set 1: The publications in co-authorship with the second SDS of the pair (specific IDR);
- Set 2: The publications in co-authorship with professors of other SDSs, but not with the second SDS of the pair (generic IDR);
- Set 3: The publications in co-authorship with professors of the same SDS (non IDR publications).

To answer our two research questions (Q1 and Q2), we verify whether the impact of publications and the prestige of the publishing journals differ significantly in the three subsets, comparing them as follows:

- Set 1 vs Set 2
- Set 1 vs Set 3
- Set (1∪2) vs Set 3

The impact of each publication and the prestige of the relative journal are measured using two indicators:

---

[4] The list of all SDSs is reported in Annex 1.
[5] We define the "specific degree of interdisciplinarity" of a field with another specific research field as the ratio between the number of publications co-authored by researchers from both fields, to the number of publications authored by researchers belonging to the first field. The reader is referred to Abramo et al. (2012) for additional details on the methodology for identifying the SDS pairs with the highest collaboration rates.

- *Article Impact Index, AII* – The ratio between the number of citations received by the publication and the average of the citations of all the national publications cited,[6] indexed in the same year and subject category.[7]
- *Journal Impact Index, JII* – The ratio between the IF of the publishing journal and the average of the IF of all the journals of the same subject category.

The publications (71,633) subject to analysis are those authored in 2004-2008 by the professors (2,279) of the first SDS of the 110 "specific" pairs. The total number of SDSs analyzed (i.e. the first SDSs of the pairs) is 72. Comparing the number of "first SDSs" to the larger number of specific pairs developed, we immediately observe that the professors of some SDSs frequently carry out IDR with more than one SDS. Of the 110 SDS pairs, the vast majority (93 (84.5%)) are composed of professors belonging to SDSs in the same UDA (i.e. the two collaborating fields belong to just one of the disciplinary areas).

In Annex 2, for each SDS pair, we report the number of publications coauthored by the professors of the first SDS in the pair, divided into the three sets considered. The total reported is that of the publications coauthored by the professors of the first SDS. The "first SDS" can participate in more than one kind of "specific" collaboration with another SDS, as well as in "generic" co-authorships and in non IDR. All of these co-authorships are counted in the same row, and it is for this that we see the different pairs (all starting with the same first SDS) listed with the same total number of publications. In detail, excluding double counting, we have 16,453 publications in set 1; 26,984 in set 2; 36,252 in set 3; and 35,381 in set (1∪2).

## 3. Results

For each subset of publications co-authored by professors belonging to the first SDS of the pairs (IDR with the second SDS; IDR with a "generic" SDS; non IDR) we measure the medians of AII and JII distributions. We then measure the differences of the medians between the subsets. After verifying that the distributions of AII and JII in each subset are not normal (Shapiro-Wilk test), we run a two-sample Wilcoxon rank-sum (Mann-Whitney) test to assess whether the differences between the distributions are significant. For each comparison, Table 1 presents the statistics for the positive and negative differences, where significant. It should be noted that when we compare the non-IDR publications of the professors of each of the 72 analyzed SDSs with the generic IDR publications (those co-authored with professors of whatever other SDS), the concept of the pair disappears: for this last column in Table 1, refers to the SDS rather than to the pair. In the comparison between set 1 and set 2, we see that 34 (31%) out of the 110 SDS pairs show significantly different AII distributions, and 43 (39%) show significantly different JII distributions. Between set 1 and set 3, the corresponding frequencies are 35 and 34. In the comparison between set 1∪2 and set 3, we see that 25 (34%) of the 72 SDSs show significantly different AII distributions and 21 (29%) significantly different JII distributions. The number of pairs/SDSs in which IDR produces products with median impact greater than the non-IDR products (i.e., set 1 vs set 3, and set 1∪2 vs set 3) is greater than the cases in which they produce lesser impact.

---

[6] This scaling factor results as the most effective at normalizing citations (Abramo et al., 2012).

[7] Per publications in multi-category journals, the value of the indicator is equal to the average of the values for the individual subject categories.

The same occurs for the products resulting from specific IDR (set 1) compared to those resulting from generic IDR (set 2). The products of specific IDR pairs are also published in journals of greater prestige for greater numbers of pairs/SDSs, compared to the products of other pairs ("generic" and non-IDR). On the other hand, the non-IDR publications suffer in comparison to the total IDR publications, being more often published in journals of lesser prestige.

|  |  | set 1 vs set 2 | set 1 vs set 3 | set(1 U 2) vs set 3 |
|---|---|---|---|---|
| Indicator | Δ median > 0 | No. of pairs | No. of pairs | No. of SDSs |
| AII | Y | 25 (22.7%) | 30 (27.3%) | 20 (27.8%) |
|  | N | 9 (8.2%) | 5 (4.5%) | 5 (6.9%) |
| JII | Y | 29 (26.4%) | 27 (24.5%) | 15 (20.8%) |
|  | N | 14 (12.7%) | 7 (6.4%) | 6 (8.3%) |

*Table 1: Number of pairs and SDSs where differences of AII and JII distributions are significant*

It is also interesting to analyze the extent of these differences, where significant. In Annex 4, for each pair of subsets compared, for all of the SDS pairs (specific vs generic IDR, specific vs non IDR; specific and generic IDR vs non IDR only), we present the differences of the medians for the indicator AII. Annex 5 provides the same calculations for the differences concerning JII medians. In Table 2, then we provide the extract of the five SDS pairs (or SDSs) with the highest differences of AII and JII medians (positive and negative). These data are highly informative. For example we observe that the IDR carried out jointly by professors of applied physical chemistry (ING-IND/23) and foundations of chemistry for technologies (CHIM/07) leads to results with normalized median impact greater (+0.81) than that of the publications resulting from professors of applied physical chemistry and colleagues of specializations other than foundations of chemistry for technologies. Vice versa, the IDR results from professors of applied geophysics (GEO/11) working with professors of solid earth geophysics (GEO/10) have a lower normalized median impact (-0.43) than IDR results from the same professors of applied geophysics with colleagues in specializations other than solid earth geophysics.

|  | set 1 vs set 2 | | set 1 vs set 3 | | set(1 ∪ 2) vs set 3 | |
|---|---|---|---|---|---|---|
|  | AII | JII | AII | JII | AII | JII |
| Δ↑ | 0.81 ING-IND/23_CHIM/07 | 0.70 FIS/04_FIS/01 | 0.82 MED/49_CHIM/03 | 0.65 BIO/08_BIO/18 | 0.33 ING-IND/09 | 0.47 AGR/04 |
|  | 0.75 MED/49_CHIM/03 | 0.69 BIO/08_BIO/18 | 0.72 ING-IND/23_CHIM/07 | 0.60 AGR/04_AGR/02 | 0.25 MED/37 | 0.46 ING-IND/05 |
|  | 0.68 ING-IND/18_ING-IND/19 | 0.58 MED/37_MED/26 | 0.47 ING-IND/18_ING-IND/19 | 0.46 FIS/04_FIS/01 | 0.24 FIS/04 | 0.41 FIS/04 |
|  | 0.54 BIO/17_MED/04 | 0.55 ICAR/01_ICAR/02 | 0.44 BIO/08_BIO/18 | 0.41 MED/46_MED/09 | 0.20 MED/12 | 0.36 VET/09 |
|  | 0.42 BIO/08_BIO/18 | 0.48 GEO/09_GEO/06 | 0.41 MED/37_MED/26 | 0.41 MED/15_MED/09 | 0.20 AGR/18 | 0.24 BIO/02 |
| Δ↓ | -0.18 MED/10_MED/09 | -0.30 MED/37_MED/27 | -0.04 FIS/03_FIS/01 | -0.31 GEO/09_GEO/07 | -0.06 MED/04 | -0.09 BIO/17 |
|  | -0.21 MED/37_MED/27 | -0.43 GEO/09_GEO/07 | -0.22 BIO/17_BIO/16 | -0.34 BIO/11_BIO/10 | -0.12 BIO/17 | -0.13 FIS/03 |
|  | -0.27 MED/21_MED/18 | -0.48 MED/49_BIO/12 | -0.28 BIO/11_BIO/10 | -0.41 MED/22_MED/36 | -0.23 BIO/11 | -0.17 CHIM/04 |
|  | -0.27 MED/46_BIO/10 | -0.56 MED/22_MED/18 | -0.29 MED/46_BIO/10 | -0.53 GEO/11_GEO/10 | -0.24 GEO/07 | -0.24 BIO/11 |
|  | -0.43 GEO/11_GEO/10 | -0.72 GEO/11_GEO/10 | -0.36 GEO/12_FIS/06 | -0.62 MED/22_MED/18 | -0.28 GEO/12 | -0.40 GEO/12 |

*Table 2: Maximum significant differences (positive and negative) between the medians of the indicators AII and JII, for the publications in the subsets compared*

Following the above general analysis, we now verify if there are specificities at the discipline level. For this, we carry out the three comparisons between the subsets of publications, with the SDS pairs (SDSs) grouped by the UDA to which the first SDS belongs. Table 3 presents the results of the comparison between set 1 and set 2 (specific IDR vs generic IDR). In all the UDAs except Earth sciences and Agricultural and veterinary sciences we observe that there are always more pairs in which $\Delta_{median}$ of AII is positive than there are pairs where it is negative (9 pairs out of a total 34). The same occurs for JII (SDS pairs with negative difference are now 13 out of 43).

Table 4 presents the results of the comparison between set 1 and set 3 (specific IDR vs non-IDR). The dominance of pairs with positive $\Delta_{median}$ remains the same, with Earth sciences again the exception, for both AII and JII. The median differences of AII are negative in 5 SDS pairs out of 35, and in 7 pairs out of 34 for JII.

Finally, Table 5 presents the values for the comparison between set (1 ∪ 2) and set 3 (all IDR vs non-IDR). For AII, the median differences are negative in 5 SDSs out of 25: Earth sciences together with Biology present more SDSs with negative difference than with positive. For JII, it is again Earth sciences the only UDA to show SDSs (2 in all) with a lower median for IDR resulting publications.

|  |  | AII | | JII | |
|---|---|---|---|---|---|
| UDA‡ | Tot. no. pairs | No. pairs* | Δ median | No. pairs* | Δ median |
| Agricultural and veterinary sciences | 11 (10.0%) | 1 (9.1%) | 0(+) \| 1(-) | 0 | - |
| Biology | 11 (10.0%) | 3 (27.3%) | 2(+) \| 1(-) | 5 (45.5%) | 3(+) \| 2(-) |
| Chemistry | 11 (10.0%) | 3 (27.3%) | 3(+) \| 0(-) | 3 (27.3%) | 3(+) \| 0(-) |
| Earth sciences | 6 (5.5%) | 1 (16.7%) | 0(+) \| 1(-) | 5 (83.3%) | 2(+) \| 3(-) |
| Civil engineering | 1 (0.9%) | 1 (100.0%) | 1(+) \| 0(-) | 1 (100.0%) | 1(+) \| 0(-) |
| Industrial and information engineering | 9 (8.2%) | 3 (33.3%) | 3(+) \| 0(-) | 4 (44.4%) | 4(+) \| 0(-) |
| Mathematics | 1 (0.9%) | 0 | - | 0 | - |
| Medicine | 56 (50.9%) | 19 (33.9%) | 13(+) \| 6(-) | 23 (41.1%) | 15(+) \| 8(-) |
| Physics | 4 (3.6%) | 3 (75.0%) | 3(+) \| (-) | 2 (50.0%) | 2(+) \| 0(-) |
| Total | 110 | 34 (30.9%) | 25(+) \| 9(-) | 43 (39.1%) | 30(+) \| 13(-) |

*Table 3: Two-sample Wilcoxon rank-sum (Mann-Whitney) test - comparison of set 1 to set 2*
*\* Number of pairs where the differences of AII and JII distributions are statistically significant*
*‡ The UDAs are those to which the first SDS of the pair belongs*

|  |  | AII | | JII | |
|---|---|---|---|---|---|
| UDA‡ | Tot. no. pairs | No. pairs* | Δ median | No. pairs* | Δ median |
| Agricultural and veterinary sciences | 11 (10.0%) | 2 (18.2%) | 2(+) \| 0(-) | 3 (27.3%) | 3(+) \| 0(-) |
| Biology | 11 (10.0%) | 4 (36.4%) | 2(+) \| 2(-) | 6 (54.5%) | 4(+) \| 2(-) |
| Chemistry | 11 (10.0%) | 2 (18.2%) | 2(+) \| 0(-) | 2 (18.2%) | 2(+) \| 0(-) |
| Earth sciences | 6 (5.5%) | 1 (16.7%) | 0(+) \| 1(-) | 3 (50.0%) | 1(+) \| 2(-) |
| Civil engineering | 1 (0.9%) | 0 | - | 0 | - |
| Industrial and information engineering | 9 (8.2%) | 3 (33.3%) | 3(+) \| 0(-) | 1 (11.1%) | 1(+) \| 0(-) |
| Mathematics | 1 (0.9%) | 0 | - | 0 | - |
| Medicine | 56 (50.9%) | 20 (35.7%) | 19(+) \| 1(-) | 16 (28.6%) | 14(+) \| 2(-) |
| Physics | 4 (3.6%) | 3 (75.0%) | 2(+) \| 1(-) | 3 (75.0%) | 2(+) \| 1(-) |
| Total | 110 | 35 (31.8%) | 30(+) \| 5(-) | 34 (30.9%) | 27(+) \| 7(-) |

*Table4: Two-sample Wilcoxon rank-sum (Mann-Whitney) test - comparison of set 1 to set 3*
*\* Number of pairs where the differences of AII and JII distributions are statistically significant*
*‡ The UDAs are those to which the first SDS of the pair belongs*

|  |  | AII | | JII | |
|---|---|---|---|---|---|
| UDA‡ | Tot. no. | No. | Δ median | No. | Δ median |

|  | first SDS | first SDS* |  | first SDS* |  |
|---|---|---|---|---|---|
| Agricultural and veterinary sciences | 8 (11.1%) | 3 (37.5%) | 3(+) \| 0(-) | 3 (37.5%) | 3(+) \| 0(-) |
| Biology | 7 (9.7%) | 2 (28.6%) | 0(+) \| 2(-) | 4 (57.1%) | 2(+) \| 2(-) |
| Chemistry | 7 (9.7%) | 0 | - | 2 (28.6%) | 1(+) \| 1(-) |
| Earth sciences | 5 (6.9%) | 2 (40.0%) | 0(+) \| 2(-) | 2 (40.0%) | 0(+) \| 2(-) |
| Civil engineering | 1 (1.4%) | 0 | - | 0 | - |
| Industrial and information engineering | 8 (11.1%) | 2 (25.0%) | 2(+) \| 0(-) | 1 (12.5%) | 1(+) \| 0(-) |
| Mathematics | 1 (1.4%) | 0 | - | 0 | - |
| Medicine | 31 (43.1%) | 14 (45.2%) | 13(+) \| 1(-) | 5 (16.1%) | 5(+) \| 0(-) |
| Physics | 4 (5.6%) | 2 (50.0%) | 2(+) \| 0(-) | 4 (100.0%) | 3(+) \| 1(-) |
| Total | 72 | 25 (34.7%) | 20(+) \| 5(-) | 21 (29.2%) | 15(+) \| 6(-) |

*Table 5: Two-sample Wilcoxon rank-sum (Mann-Whitney) test - comparison of set (1∪2) to set 3*
*\* Number of first SDS where the differences of AII and JII distributions are statistically significant*
*‡ The UDAs are those to which SDS1 belongs*

## 4. Conclusions

The question of interest is whether interdisciplinary research teams achieve knowledge gains of greater impact. The answer is mixed.

The occurrence of frequent collaborations between specialists in a pair of research sectors could signal the emergence of a new field, which initially has important connotations to the two forbearer fields. However, such connotations seem inevitably bound to diffuse and decline. Through the analysis of the scientific sectors of co-authors and the interdisciplinary collaborations involved, we have succeeded in identifying a set of publications featuring recurring collaboration between specific pairs of sectors, which we call "specific interdisciplinary research". We distinguish these from the publications resulting from what we call "generic interdisciplinary research", meaning from teams involving sectors where collaborations are less frequent. We then compared the impact and prestige of the publishing journals: i) for the "specific" versus the "generic" publications; ii) for the "specific" versus the non-interdisciplinary publications; iii) of all the interdisciplinary versus the non-interdisciplinary publications. The comparisons show significant differences in a third or slightly more of the cases, varying somewhat with the sets compared and the indicator. In general, specific interdisciplinary research delivers more cases of greater gains to science than generic IDR and non IDR; overall IDR delivers more cases of greater gains than non IDR. This holds true in all disciplines except for Earth sciences. While these data are indicative, they do not permit a definitive response to the research questions.

An interesting aspect of the study in hand is that we are able to observe which interdisciplinary combinations pay off in results, and which do not. Still more interesting would be to understand why such differences occur. In fact, while it is intuitive that interdisciplinary outputs would be cited in broad sets of fields, and therefore gain more citations, it seems more difficult to understand a "negative" result from interdisciplinarity. Perhaps the publications co-authored by researchers from very distant fields experience delayed recognition and are highly cited only in the long run, thus not showing up in a short citation window. It has also been noted that publications from "distant" collaborations are typically published in journals with a lower than expected impact factor (Wang et al., 2016). Thus, for these specific subsets of publications our study may have suffered from an evaluation bias. The concerns and questions raised by these aspects of our study could be the subject of future investigation.

# Annex 1 – SDS list

| Code | Title | UDA |
|---|---|---|
| MAT/01 | Mathematical Logic | Mathematics and computer sciences |
| MAT/02 | Algebra | Mathematics and computer sciences |
| MAT/03 | Geometry | Mathematics and computer sciences |
| MAT/04 | Complementary Mathematics | Mathematics and computer sciences |
| MAT/05 | Mathematical Analysis | Mathematics and computer sciences |
| MAT/06 | Probability and Mathematical Statistics | Mathematics and computer sciences |
| MAT/07 | Mathematical Physics | Mathematics and computer sciences |
| MAT/08 | Numerical Analysis | Mathematics and computer sciences |
| MAT/09 | Operational Research | Mathematics and computer sciences |
| INF/01 | Computer Science | Mathematics and computer sciences |
| FIS/01 | Experimental Physics | Physics |
| FIS/02 | Theoretical Physics, Mathematical Models and Methods | Physics |
| FIS/03 | Physics of Matter | Physics |
| FIS/04 | Nuclear and Subnuclear Physics | Physics |
| FIS/05 | Astronomy and Astrophysics | Physics |
| FIS/06 | Physics for Earth and Atmospheric Sciences | Physics |
| FIS/07 | Applied Physics (Cultural Heritage, Environment, Biology …) | Physics |
| FIS/08 | Didactics and History of Physics | Physics |
| CHIM/01 | Analytical Chemistry | Chemistry |
| CHIM/02 | Physical Chemistry | Chemistry |
| CHIM/03 | General and Inorganic Chemistry | Chemistry |
| CHIM/04 | Industrial Chemistry | Chemistry |
| CHIM/05 | Science and Technology of Polymeric Materials | Chemistry |
| CHIM/06 | Organic Chemistry | Chemistry |
| CHIM/07 | Foundations of Chemistry for Technologies | Chemistry |
| CHIM/08 | Pharmaceutical Chemistry | Chemistry |
| CHIM/09 | Applied Technological Pharmaceutics | Chemistry |
| CHIM/10 | Food Chemistry | Chemistry |
| CHIM/11 | Chemistry and Biotechnology of Fermentations | Chemistry |
| CHIM/12 | Environmental Chemistry and Chemistry for Cultural Heritage | Chemistry |
| GEO/01 | Palaeontology and Palaeoecology | Earth sciences |
| GEO/02 | Stratigraphic and Sedimentological Geology | Earth sciences |
| GEO/03 | Structural Geology | Earth sciences |
| GEO/04 | Physical Geography and Geomorphology | Earth sciences |
| GEO/05 | Applied Geology | Earth sciences |
| GEO/06 | Mineralogy | Earth sciences |
| GEO/07 | Petrology and Petrography | Earth sciences |
| GEO/08 | Geochemistry and Volcanology | Earth sciences |
| GEO/09 | Mineral Geological Resources and Mineralogic and Petrographic Applications for the Environment and Cultural Heritage | Earth sciences |
| GEO/10 | Solid Earth Geophysics | Earth sciences |
| GEO/11 | Applied Geophysics | Earth sciences |
| GEO/12 | Oceanography and Atmospheric Physics | Earth sciences |
| BIO/01 | General Botanics | Biology |
| BIO/02 | Systematic Botanics | Biology |
| BIO/03 | Environmental and Applied Botanics | Biology |
| BIO/04 | Vegetal Physiology | Biology |
| BIO/05 | Zoology | Biology |
| BIO/06 | Comparative Anatomy and Citology | Biology |
| BIO/07 | Ecology | Biology |
| BIO/08 | Anthropology | Biology |
| BIO/09 | Physiology | Biology |
| BIO/10 | Biochemistry | Biology |
| BIO/11 | Molecular Biology | Biology |
| BIO/12 | Clinical Biochemistry and Biology | Biology |
| BIO/13 | Applied Biology | Biology |
| BIO/14 | Pharmacology | Biology |
| BIO/15 | Pharmaceutic Biology | Biology |

| Code | Title | UDA |
|---|---|---|
| BIO/16 | Human Anatomy | Biology |
| BIO/17 | Histology | Biology |
| BIO/18 | Genetics | Biology |
| BIO/19 | General Microbiology | Biology |
| MED/01 | Medical Statistics | Medicine |
| MED/02 | History of Medicine | Medicine |
| MED/03 | Medical Genetics | Medicine |
| MED/04 | General Pathology | Medicine |
| MED/05 | Clinical Pathology | Medicine |
| MED/06 | Medical Oncology | Medicine |
| MED/07 | Microbiology and Clinical Microbiology | Medicine |
| MED/08 | Pathological Anatomy | Medicine |
| MED/09 | Internal Medicine | Medicine |
| MED/10 | Respiratory Diseases | Medicine |
| MED/11 | Cardiovascular Diseases | Medicine |
| MED/12 | Gastroenterology | Medicine |
| MED/13 | Endocrinology | Medicine |
| MED/14 | Nephrology | Medicine |
| MED/15 | Blood Diseases | Medicine |
| MED/16 | Rheumatology | Medicine |
| MED/17 | Infectious Diseases | Medicine |
| MED/18 | General Surgery | Medicine |
| MED/19 | Plastic Surgery | Medicine |
| MED/20 | Pediatric and Infant Surgery | Medicine |
| MED/21 | Thoracic Surgery | Medicine |
| MED/22 | Vascular Surgery | Medicine |
| MED/23 | Cardiac Surgery | Medicine |
| MED/24 | Urology | Medicine |
| MED/25 | Psychiatry | Medicine |
| MED/26 | Neurology | Medicine |
| MED/27 | Neurosurgery | Medicine |
| MED/28 | Odonto-Stomalogical Diseases | Medicine |
| MED/29 | Maxillofacial Surgery | Medicine |
| MED/30 | Eye Diseases | Medicine |
| MED/31 | Otorinolaringology | Medicine |
| MED/32 | Audiology | Medicine |
| MED/33 | Locomotory Diseases | Medicine |
| MED/34 | Physical and Rehabilitation Medicine | Medicine |
| MED/35 | Skin and Venereal Diseases | Medicine |
| MED/36 | Diagnostic Imaging and Radiotherapy | Medicine |
| MED/37 | Neuroradiology | Medicine |
| MED/38 | General and Specialised Pediatrics | Medicine |
| MED/39 | Child Neuropsychiatry | Medicine |
| MED/40 | Gynaecology and Obstetrics | Medicine |
| MED/41 | Anaesthesiology | Medicine |
| MED/42 | General and Applied Hygiene | Medicine |
| MED/43 | Legal Medicine | Medicine |
| MED/44 | Occupational Medicine | Medicine |
| MED/45 | General, Clinical and Pediatric Nursing | Medicine |
| MED/46 | Laboratory Medicine Techniques | Medicine |
| MED/47 | Nursing and Midwifery | Medicine |
| MED/48 | Neuropsychiatric and Rehabilitation Nursing | Medicine |
| MED/49 | Applied Dietary Sciences | Medicine |
| MED/50 | Applied Medical Sciences | Medicine |
| AGR/01 | Rural Economy and Evaluation | Agricultural and veterinary sciences |
| AGR/02 | Agronomy and Herbaceous Cultivation | Agricultural and veterinary sciences |
| AGR/03 | General Arboriculture and Tree Cultivation | Agricultural and veterinary sciences |
| AGR/04 | Horticulture and Floriculture | Agricultural and veterinary sciences |
| AGR/05 | Forestry and Silviculture | Agricultural and veterinary sciences |
| AGR/06 | Wood Technology and Woodland Management | Agricultural and veterinary sciences |
| AGR/07 | Agrarian Genetics | Agricultural and veterinary sciences |

| Code | Title | UDA |
|---|---|---|
| AGR/08 | Agrarian Hydraulics and Hydraulic Forest Management | Agricultural and veterinary sciences |
| AGR/09 | Agricultural Mechanics | Agricultural and veterinary sciences |
| AGR/10 | Rural Construction and Environmental Land Management | Agricultural and veterinary sciences |
| AGR/11 | General and Applied Entomology | Agricultural and veterinary sciences |
| AGR/12 | Plant Pathology | Agricultural and veterinary sciences |
| AGR/13 | Agricultural Chemistry | Agricultural and veterinary sciences |
| AGR/14 | Pedology | Agricultural and veterinary sciences |
| AGR/15 | Food Sciences | Agricultural and veterinary sciences |
| AGR/16 | Agricultural Microbiology | Agricultural and veterinary sciences |
| AGR/17 | General Techniques for Zoology and Genetic Improvement | Agricultural and veterinary sciences |
| AGR/18 | Animal Nutrition and Feeding | Agricultural and veterinary sciences |
| AGR/19 | Special Techniques for Zoology | Agricultural and veterinary sciences |
| AGR/20 | Animal Husbandry | Agricultural and veterinary sciences |
| VET/01 | Anatomy of Domestic Animals | Agricultural and veterinary sciences |
| VET/02 | Veterinary Physiology | Agricultural and veterinary sciences |
| VET/03 | General Pathology and Veterinary Pathological Anatomy | Agricultural and veterinary sciences |
| VET/04 | Inspection of Food Products of Animal Origin | Agricultural and veterinary sciences |
| VET/05 | Infectious Diseases of Domestic Animals | Agricultural and veterinary sciences |
| VET/06 | Parasitology and Parasitic Animal Diseases | Agricultural and veterinary sciences |
| VET/07 | Veterinary Pharmacology and Toxicology | Agricultural and veterinary sciences |
| VET/08 | Clinical Veterinary Medicine | Agricultural and veterinary sciences |
| VET/09 | Clinical Veterinary Surgery | Agricultural and veterinary sciences |
| VET/10 | Clinical Veterinary Obstetrics and Gynaecology | Agricultural and veterinary sciences |
| ICAR/01 | Hydraulics | Civil engineering |
| ICAR/02 | Maritime Hydraulic Construction and Hydrology | Civil engineering |
| ICAR/03 | Environmental and Health Engineering | Civil engineering |
| ICAR/04 | Road, Railway and Airport Construction | Civil engineering |
| ICAR/05 | Transport | Civil engineering |
| ICAR/06 | Topography and Cartography | Civil engineering |
| ICAR/07 | Geotechnics | Civil engineering |
| ICAR/08 | Construction Science | Civil engineering |
| ICAR/09 | Construction Techniques | Civil engineering |
| ICAR/10 | Technical Architecture | Civil engineering |
| ICAR/11 | Building Production | Civil engineering |
| ICAR/12 | Architecture Technology | Civil engineering |
| ICAR/13 | Industrial Design | Civil engineering |
| ICAR/14 | Architectural and Urban Composition | Civil engineering |
| ICAR/15 | Landscape Architecture | Civil engineering |
| ICAR/16 | Interior Architecture and Venue Design | Civil engineering |
| ICAR/17 | Design | Civil engineering |
| ICAR/18 | History of Architecture | Civil engineering |
| ICAR/19 | Restoration | Civil engineering |
| ICAR/20 | Urban Planning | Civil engineering |
| ICAR/21 | Urban Studies | Civil engineering |
| ICAR/22 | Cadastral Surveying | Civil engineering |
| ING-IND/01 | Naval Architecture | Industrial and information engineering |
| ING-IND/02 | Naval and Marine Construction and Installation | Industrial and information engineering |
| ING-IND/03 | Flight Mechanics | Industrial and information engineering |
| ING-IND/04 | Aerospace Construction and Installation | Industrial and information engineering |
| ING-IND/05 | Aerospace Systems | Industrial and information engineering |
| ING-IND/06 | Fluid Dynamics | Industrial and information engineering |
| ING-IND/07 | Aerospatial Propulsion | Industrial and information engineering |
| ING-IND/08 | Fluid Machines | Industrial and information engineering |
| ING-IND/09 | Energy and Environmental Systems | Industrial and information engineering |
| ING-IND/10 | Technical Physics | Industrial and information engineering |
| ING-IND/11 | Environmental Technical Physics | Industrial and information engineering |
| ING-IND/12 | Mechanical and Thermal Measuring Systems | Industrial and information engineering |
| ING-IND/13 | Applied Mechanics for Machinery | Industrial and information engineering |
| ING-IND/14 | Mechanical Design and Machine Building | Industrial and information engineering |
| ING-IND/15 | Design and Methods for Industrial Engineering | Industrial and information engineering |
| ING-IND/16 | Production Technologies and Systems | Industrial and information engineering |

| Code | Title | UDA |
|---|---|---|
| ING-IND/17 | Industrial and Mechanical Plant | Industrial and information engineering |
| ING-IND/18 | Nuclear Reactor Physics | Industrial and information engineering |
| ING-IND/19 | Nuclear Plants | Industrial and information engineering |
| ING-IND/20 | Nuclear Measurement Tools | Industrial and information engineering |
| ING-IND/21 | Metallurgy | Industrial and information engineering |
| ING-IND/22 | Science and Technology of Materials | Industrial and information engineering |
| ING-IND/23 | Applied Physical Chemistry | Industrial and information engineering |
| ING-IND/24 | Principles of Chemical Engineering | Industrial and information engineering |
| ING-IND/25 | Chemical Plants | Industrial and information engineering |
| ING-IND/26 | Theory of Development for Chemical Processes | Industrial and information engineering |
| ING-IND/27 | Industrial and Technological Chemistry | Industrial and information engineering |
| ING-IND/28 | Excavation Engineering and Safety | Industrial and information engineering |
| ING-IND/29 | Raw Materials Engineering | Industrial and information engineering |
| ING-IND/30 | Hydrocarburants and Fluids of the Subsoil | Industrial and information engineering |
| ING-IND/31 | Electrotechnics | Industrial and information engineering |
| ING-IND/32 | Electrical Convertors, Machines and Switches | Industrial and information engineering |
| ING-IND/33 | Electrical Energy Systems | Industrial and information engineering |
| ING-IND/34 | Industrial Bioengineering | Industrial and information engineering |
| ING-IND/35 | Engineering and Management | Industrial and information engineering |
| ING-INF/01 | Electronics | Industrial and information engineering |
| ING-INF/02 | Electromagnetic Fields | Industrial and information engineering |
| ING-INF/03 | Telecommunications | Industrial and information engineering |
| ING-INF/04 | Automatics | Industrial and information engineering |
| ING-INF/05 | Data Processing Systems | Industrial and information engineering |
| ING-INF/06 | Electronic and Information Bioengineering | Industrial and information engineering |
| ING-INF/07 | Electric and Electronic Measurement Systems | Industrial and information engineering |

**Annex 2 - SDS pairs with specific degree of interdisciplinarity greater than 10%**
*Data 2004-2008 for SDSs with at least 100 publications*

| SDS | Pairs | set 1 | set 2 | set 3 | (set 1 ∪ 2) | Total |
|---|---|---|---|---|---|---|
| AGR/04 | AGR/04_AGR/02 | 38 (13.5%) | 60 (21.4%) | 183 (65.1%) | 98 (34.9%) | 281 (100%) |
| AGR/17 | AGR/17_AGR/19 | 159 (30.6%) | 116 (22.3%) | 245 (47.1%) | 275 (52.9%) | 520 (100%) |
| AGR/18 | AGR/18_AGR/17; AGR/18_AGR/19 | 206 (20.2%) | 432 (42.4%) | 382 (37.5%) | 638 (62.5%) | 1,020 (100%) |
| AGR/20 | AGR/20_AGR/18; AGR/20_AGR/19 | 73 (13.6%) | 259 (48.3%) | 204 (38.1%) | 332 (61.9%) | 536 (100%) |
| BIO/02 | BIO/02_BIO/03 | 54 (14.2%) | 155 (40.7%) | 172 (45.1%) | 209 (54.9%) | 381 (100%) |
| BIO/08 | BIO/08_BIO/18 | 31 (11.7%) | 72 (27.1%) | 163 (61.3%) | 103 (38.7%) | 266 (100%) |
| BIO/11 | BIO/11_BIO/10 | 397 (23.2%) | 705 (41.3%) | 606 (35.5%) | 1,102 (64.5%) | 1,708 (100%) |
| BIO/12 | BIO/12_BIO/10; BIO/12_MED/09 | 677 (18.7%) | 2,001 (55.2%) | 946 (26.1%) | 2,678 (73.9%) | 3,624 (100%) |
| BIO/15 | BIO/15_BIO/14; BIO/15_CHIM/06 | 203 (13.6%) | 733 (49.0%) | 560 (37.4%) | 936 (62.6%) | 1,496 (100%) |
| BIO/17 | BIO/17_BIO/16; BIO/17_MED/04 | 426 (14.9%) | 1,736 (60.6%) | 702 (24.5%) | 2,162 (75.5%) | 2,864 (100%) |
| BIO/19 | BIO/19_BIO/10; BIO/19_MED/07 | 143 (11.8%) | 723 (59.6%) | 348 (28.7%) | 866 (71.3%) | 1,214 (100%) |
| CHIM/02 | CHIM/02_CHIM/03 | 687 (12.9%) | 1,632 (30.6%) | 3,012 (56.5%) | 2,319 (43.5%) | 5,331 (100%) |
| CHIM/04 | CHIM/04_CHIM/02; CHIM/04_CHIM/03 | 327 (10.8%) | 1,175 (39.0%) | 1,514 (50.2%) | 1,502 (49.8%) | 3,016 (100%) |
| CHIM/07 | CHIM/07_CHIM/03 | 337 (15.2%) | 1,060 (47.7%) | 827 (37.2%) | 1,397 (62.8%) | 2,224 (100%) |
| CHIM/09 | CHIM/09_CHIM/08 | 157 (12.8%) | 546 (44.5%) | 525 (42.8%) | 703 (57.2%) | 1,228 (100%) |
| CHIM/10 | CHIM/10_BIO/14; CHIM/10_CHIM/01 CHIM/10_CHIM/06 | 163 (10.4%) | 899 (57.6%) | 498 (31.9%) | 1,062 (68.1%) | 1,560 (100%) |
| CHIM/11 | CHIM/11_BIO/10 | 32 (9.0%) | 181 (50.8%) | 143 (40.2%) | 213 (59.8%) | 356 (100%) |
| CHIM/12 | CHIM/12_CHIM/01; CHIM/12_CHIM/02 | 211 (19.9%) | 473 (44.7%) | 374 (35.3%) | 684 (64.7%) | 1,058 (100%) |
| FIS/03 | FIS/03_FIS/01 | 1,814 (26.6%) | 1,094 (16.0%) | 3,919 (57.4%) | 2,908 (42.6%) | 6,827 (100%) |
| FIS/04 | FIS/04_FIS/01 | 1,549 (61.7%) | 164 (6.5%) | 796 (31.7%) | 1,713 (68.3%) | 2,509 (100%) |
| FIS/06 | FIS/06_FIS/01 | 46 (12.5%) | 116 (31.5%) | 206 (56.0%) | 162 (44.0%) | 368 (100%) |
| FIS/07 | FIS/07_FIS/01 | 826 (27.7%) | 1,114 (37.3%) | 1,046 (35.0%) | 1,940 (65.0%) | 2,986 (100%) |
| GEO/01 | GEO/01_GEO/02 | 117 (21.3%) | 91 (16.5%) | 342 (62.2%) | 208 (37.8%) | 550 (100%) |
| GEO/07 | GEO/07_GEO/08 | 76 (11.8%) | 234 (36.3%) | 334 (51.9%) | 310 (48.1%) | 644 (100%) |
| GEO/09 | GEO/09_GEO/06; GEO/09_GEO/07 | 116 (17.2%) | 282 (41.7%) | 278 (41.1%) | 398 (58.9%) | 676 (100%) |
| GEO/11 | GEO/11_GEO/10 | 31 (12.8%) | 72 (29.8%) | 139 (57.4%) | 103 (42.6%) | 242 (100%) |
| GEO/12 | GEO/12_FIS/06 | 11 (7.9%) | 27 (19.4%) | 101 (72.7%) | 38 (27.3%) | 139 (100%) |
| ICAR/01 | ICAR/01_ICAR/02 | 57 (10.8%) | 73 (13.8%) | 400 (75.5%) | 130 (24.5%) | 530 (100%) |
| ING-IND/05 | ING-IND/05_ING-IND/04 | 9 (6.9%) | 24 (18.3%) | 98 (74.8%) | 33 (25.2%) | 131 (100%) |
| ING-IND/09 | ING-IND/09_ING-IND/08 | 89 (30.8%) | 49 (17.0%) | 151 (52.2%) | 138 (47.8%) | 289 (100%) |
| ING-IND/11 | ING-IND/11_ING-IND/10 | 73 (23.6%) | 43 (13.9%) | 193 (62.5%) | 116 (37.5%) | 309 (100%) |
| ING-IND/18 | ING-IND/18_ING-IND/19 | 31 (22.0%) | 18 (12.8%) | 92 (65.2%) | 49 (34.8%) | 141 (100%) |
| ING-IND/22 | ING-IND/22_CHIM/07 | 306 (15.1%) | 633 (31.2%) | 1,087 (53.7%) | 939 (46.3%) | 2,026 (100%) |
| ING-IND/23 | ING-IND/23_CHIM/07 | 41 (11.8%) | 133 (38.4%) | 172 (49.7%) | 174 (50.3%) | 346 (100%) |
| ING-IND/27 | ING-IND/27_CHIM/07; ING-IND/27_ING-IND/25 | 150 (13.8%) | 372 (34.3%) | 564 (51.9%) | 522 (48.1%) | 1,086 (100%) |
| ING-INF/07 | ING-INF/07_ING-INF/01 | 248 (21.3%) | 224 (19.2%) | 695 (59.6%) | 472 (40.4%) | 1,167 (100%) |
| MAT/01 | MAT/01_INF/01 | 22 (12.4%) | 12 (6.7%) | 144 (80.9%) | 34 (19.1%) | 178 (100%) |
| MED/01 | MED/01_MED/09 | 207 (13.4%) | 735 (47.7%) | 600 (38.9%) | 942 (61.1%) | 1,542 (100%) |
| MED/03 | MED/03_MED/38 | 193 (12.3%) | 840 (53.5%) | 537 (34.2%) | 1,033 (65.8%) | 1,570 (100%) |
| MED/04 | MED/04_MED/09 | 488 (10.2%) | 2,590 (54.3%) | 1,692 (35.5%) | 3,078 (64.5%) | 4,770 (100%) |
| MED/05 | MED/05_MED/04; MED/05_MED/09; MED/05_MED/13 | 512 (19.2%) | 1,825 (68.6%) | 324 (12.2%) | 2,337 (87.8%) | 2,661 (100%) |
| MED/06 | MED/06_MED/04; MED/06_MED/08; MED/06_MED/09; MED/06_MED/18 | 708 (12.9%) | 2,860 (52.1%) | 1,920 (35.0%) | 3,568 (65.0%) | 5,488 (100%) |
| MED/08 | MED/08_MED/18 | 619 (13.9%) | 2,736 (61.4%) | 1,104 (24.8%) | 3,355 (75.2%) | 4,459 (100%) |
| MED/10 | MED/10_MED/09 | 114 (10.7%) | 414 (38.9%) | 536 (50.4%) | 528 (49.6%) | 1,064 (100%) |
| MED/11 | MED/11_MED/09 | 326 (12.9%) | 829 (32.8%) | 1,370 (54.3%) | 1,155 (45.7%) | 2,525 (100%) |

| SDS | Pairs | set 1 | set 2 | set 3 | (set 1 ∪ 2) | Total |
|---|---|---|---|---|---|---|
| MED/12 | MED/12_MED/08; MED/12_MED/09; MED/12_MED/18 | 821 (14.8%) | 2,692 (48.4%) | 2,052 (36.9%) | 3,513 (63.1%) | 5,565 (100%) |
| MED/13 | MED/13_MED/09 | 664 (22.4%) | 1,288 (43.5%) | 1,008 (34.1%) | 1,952 (65.9%) | 2,960 (100%) |
| MED/14 | MED/14_MED/09 | 245 (21.6%) | 476 (42.0%) | 413 (36.4%) | 721 (63.6%) | 1,134 (100%) |
| MED/15 | MED/15_MED/08; MED/15_MED/09 | 532 (12.5%) | 1,924 (45.3%) | 1,788 (42.1%) | 2,456 (57.9%) | 4,244 (100%) |
| MED/16 | MED/16_MED/09 | 264 (22.8%) | 321 (27.8%) | 571 (49.4%) | 585 (50.6%) | 1,156 (100%) |
| MED/17 | MED/17_MED/07 | 182 (11.9%) | 578 (37.7%) | 775 (50.5%) | 760 (49.5%) | 1,535 (100%) |
| MED/18 | MED/18_MED/09 | 505 (11.2%) | 2,173 (48.1%) | 1,837 (40.7%) | 2,678 (59.3%) | 4,515 (100%) |
| MED/20 | MED/20_MED/38 | 79 (25.9%) | 113 (37.0%) | 113 (37.0%) | 192 (63.0%) | 305 (100%) |
| MED/21 | MED/21_MED/08; MED/21_MED/18 | 124 (16.5%) | 340 (45.2%) | 288 (38.3%) | 464 (61.7%) | 752 (100%) |
| MED/22 | MED/22_MED/18; MED/22_MED/36 | 117 (13.2%) | 399 (45.0%) | 370 (41.8%) | 516 (58.2%) | 886 (100%) |
| MED/23 | MED/23_MED/11 | 125 (14.5%) | 288 (33.3%) | 451 (52.2%) | 413 (47.8%) | 864 (100%) |
| MED/24 | MED/24_MED/08 | 131 (11.3%) | 413 (35.6%) | 617 (53.1%) | 544 (46.9%) | 1,161 (100%) |
| MED/27 | MED/27_MED/08; MED/27_MED/26 | 184 (10.6%) | 810 (46.7%) | 740 (42.7%) | 994 (57.3%) | 1,734 (100%) |
| MED/29 | MED/29_BIO/17; MED/29_MED/08; MED/29_MED/28 | 315 (21.7%) | 645 (44.4%) | 492 (33.9%) | 960 (66.1%) | 1,452 (100%) |
| MED/32 | MED/32_MED/31 | 111 (50.5%) | 61 (27.7%) | 48 (21.8%) | 172 (78.2%) | 220 (100%) |
| MED/34 | MED/34_BIO/09; MED/34_MED/26 | 53 (20.5%) | 149 (57.8%) | 56 (21.7%) | 202 (78.3%) | 258 (100%) |
| MED/35 | MED/35_MED/08 | 180 (11.5%) | 513 (32.8%) | 872 (55.7%) | 693 (44.3%) | 1,565 (100%) |
| MED/36 | MED/36_MED/18 | 343 (12.3%) | 1,444 (51.8%) | 999 (35.9%) | 1,787 (64.1%) | 2,786 (100%) |
| MED/37 | MED/37_MED/26; MED/37_MED/27; MED/37_MED/36 | 237 (24.5%) | 546 (56.5%) | 183 (18.9%) | 783 (81.1%) | 966 (100%) |
| MED/39 | MED/39_MED/26; MED/39_MED/38 | 204 (15.7%) | 552 (42.6%) | 540 (41.7%) | 756 (58.3%) | 1,296 (100%) |
| MED/46 | MED/46_BIO/10; MED/46_MED/04; MED/46_MED/09; MED/46_MED/13 | 282 (18.4%) | 1,122 (73.0%) | 132 (8.6%) | 1,404 (91.4%) | 1,536 (100%) |
| MED/49 | MED/49_BIO/10; MED/49_BIO/12; MED/49_CHIM/03; MED/49_MED/09 | 149 (18.8%) | 571 (72.1%) | 72 (9.1%) | 720 (90.9%) | 792 (100%) |
| MED/50 | MED/50_MED/09; MED/50_MED/28; MED/50_MED/36 | 87 (14.1%) | 456 (74.1%) | 72 (11.7%) | 543 (88.3%) | 615 (100%) |
| VET/07 | VET/07_BIO/14 | 25 (13.3%) | 97 (51.6%) | 66 (35.1%) | 122 (64.9%) | 188 (100%) |
| VET/08 | VET/08_VET/03 | 69 (16.9%) | 210 (51.3%) | 130 (31.8%) | 279 (68.2%) | 409 (100%) |
| VET/09 | VET/09_VET/03; VET/09_VET/08 | 83 (21.8%) | 177 (46.6%) | 120 (31.6%) | 260 (68.4%) | 380 (100%) |
| VET/10 | VET/10_VET/03 | 24 (9.6%) | 130 (51.8%) | 97 (38.6%) | 154 (61.4%) | 251 (100%) |
| | Total | 19,235 (17.0%) | 49,050 (43.2%) | 45,146 (39.8%) | 68,285 (60.2%) | 113,431 (100%) |
| | Total without duplicates | 16,453 | 26,984 | 36,252 | 35,381 | 71,633 |

**Annex 3 - Differences between AII medians for compared subsets (only SDS pairs where AII distributions are significantly different)**

| set 1 vs set 2 | | set 1 vs set 3 | | set(1∪2) vs set 3 | |
|---|---|---|---|---|---|
| Δ AII | SDS pair | Δ AII | SDS pair | Δ AII | SDS |
| 0.42 | BIO/08_BIO/18 | 0.18 | AGR/17_AGR/19 | 0.14 | AGR/17 |
| -0.13 | BIO/17_BIO/16 | 0.18 | AGR/18_AGR/19 | 0.20 | AGR/18 |
| 0.54 | BIO/17_MED/04 | 0.44 | BIO/08_BIO/18 | -0.23 | BIO/11 |
| 0.10 | CHIM/02_CHIM/03 | -0.28 | BIO/11_BIO/10 | -0.12 | BIO/17 |
| 0.21 | CHIM/04_CHIM/03 | -0.22 | BIO/17_BIO/16 | 0.24 | FIS/04 |
| 0.36 | CHIM/10_CHIM/01 | 0.37 | BIO/17_MED/04 | 0.11 | FIS/07 |
| 0.14 | FIS/04_FIS/01 | 0.07 | CHIM/02_CHIM/03 | -0.24 | GEO/07 |
| 0.28 | FIS/06_FIS/01 | 0.35 | CHIM/10_CHIM/01 | -0.28 | GEO/12 |
| 0.13 | FIS/07_FIS/01 | -0.04 | FIS/03_FIS/01 | 0.33 | ING-IND/09 |
| -0.43 | GEO/11_GEO/10 | 0.25 | FIS/04_FIS/01 | 0.04 | ING-INF/07 |
| 0.37 | ICAR/01_ICAR/02 | 0.18 | FIS/07_FIS/01 | 0.10 | MED/01 |
| 0.68 | ING-IND/18_ING-IND/19 | -0.36 | GEO/12_FIS/06 | -0.06 | MED/04 |
| 0.81 | ING-IND/23_CHIM/07 | 0.47 | ING-IND/18_ING-IND/19 | 0.13 | MED/06 |
| 0.36 | ING-IND/27_CHIM/07 | 0.72 | ING-IND/23_CHIM/07 | 0.04 | MED/10 |
| 0.11 | MED/01_MED/09 | 0.26 | ING-IND/27_CHIM/07 | 0.10 | MED/11 |
| 0.13 | MED/05_MED/13 | 0.20 | MED/01_MED/09 | 0.20 | MED/12 |
| 0.13 | MED/06_MED/08 | 0.22 | MED/06_MED/08 | 0.14 | MED/14 |
| -0.18 | MED/10_MED/09 | 0.24 | MED/06_MED/09 | 0.16 | MED/15 |
| -0.10 | MED/12_MED/18 | 0.16 | MED/11_MED/09 | 0.15 | MED/17 |
| 0.06 | MED/13_MED/09 | 0.20 | MED/12_MED/08 | 0.13 | MED/18 |
| 0.18 | MED/15_MED/09 | 0.27 | MED/12_MED/09 | 0.16 | MED/21 |
| 0.06 | MED/18_MED/09 | 0.11 | MED/12_MED/18 | 0.15 | MED/22 |
| 0.23 | MED/21_MED/08 | 0.13 | MED/14_MED/09 | 0.18 | MED/23 |
| -0.27 | MED/21_MED/18 | 0.15 | MED/15_MED/08 | 0.25 | MED/37 |
| 0.17 | MED/29_BIO/17 | 0.30 | MED/15_MED/09 | 0.19 | VET/09 |
| -0.17 | MED/29_MED/08 | 0.16 | MED/17_MED/07 | | |
| 0.13 | MED/29_MED/28 | 0.18 | MED/18_MED/09 | | |
| 0.27 | MED/37_MED/26 | 0.27 | MED/21_MED/08 | | |
| -0.21 | MED/37_MED/27 | 0.19 | MED/23_MED/11 | | |
| 0.20 | MED/39_MED/26 | 0.15 | MED/27_MED/26 | | |
| -0.27 | MED/46_BIO/10 | 0.17 | MED/29_BIO/17 | | |
| 0.41 | MED/49_BIO/10 | 0.41 | MED/37_MED/26 | | |
| 0.75 | MED/49_CHIM/03 | 0.14 | MED/39_MED/26 | | |
| -0.17 | VET/08_VET/03 | -0.29 | MED/46_BIO/10 | | |
| | | 0.82 | MED/49_CHIM/03 | | |

**Annex 4 - Differences between JII medians for compared subsets (only SDSs pairs where JII distributions are significantly different)**

| set 1 vs set 2 | | set 1 vs set 3 | | set(1∪2) vs set 3 | |
|---|---|---|---|---|---|
| Δ JII | SDS pair | Δ JII | SDS pair | Δ JII | SDS |
| 0.69 | BIO/08_BIO/18 | 0.60 | AGR/04_AGR/02 | 0.47 | AGR/04 |
| -0.15 | BIO/11_BIO/10 | 0.65 | BIO/08_BIO/18 | 0.21 | AGR/18 |
| -0.23 | BIO/17_BIO/16 | -0.34 | BIO/11_BIO/10 | 0.24 | BIO/02 |
| 0.44 | BIO/17_MED/04 | 0.24 | BIO/15_BIO/14 | -0.24 | BIO/11 |
| 0.26 | BIO/19_MED/07 | -0.27 | BIO/17_BIO/16 | 0.19 | BIO/15 |
| 0.19 | CHIM/04_CHIM/02 | 0.28 | BIO/17_MED/04 | -0.09 | BIO/17 |
| 0.35 | CHIM/04_CHIM/03 | 0.21 | BIO/19_MED/07 | -0.17 | CHIM/04 |
| 0.13 | CHIM/07_CHIM/03 | 0.13 | CHIM/07_CHIM/03 | 0.14 | CHIM/09 |
| 0.70 | FIS/04_FIS/01 | 0.13 | CHIM/09_CHIM/08 | -0.13 | FIS/03 |
| 0.07 | FIS/07_FIS/01 | -0.14 | FIS/03_FIS/01 | 0.41 | FIS/04 |
| -0.17 | GEO/01_GEO/02 | 0.46 | FIS/04_FIS/01 | 0.22 | FIS/06 |
| 0.28 | GEO/07_GEO/08 | 0.16 | FIS/07_FIS/01 | 0.13 | FIS/07 |
| 0.48 | GEO/09_GEO/06 | 0.25 | GEO/09_GEO/06 | -0.09 | GEO/07 |
| -0.43 | GEO/09_GEO/07 | -0.31 | GEO/09_GEO/07 | -0.40 | GEO/12 |
| -0.72 | GEO/11_GEO/10 | -0.53 | GEO/11_GEO/10 | 0.46 | ING-IND/05 |
| 0.55 | ICAR/01_ICAR/02 | 0.11 | ING-INF/07_ING-INF/01 | 0.15 | MED/18 |
| 0.28 | ING-IND/18_ING-IND/19 | 0.19 | MED/04_MED/09 | 0.16 | MED/23 |
| 0.20 | ING-IND/22_CHIM/07 | 0.11 | MED/05_MED/04 | 0.06 | MED/27 |
| 0.27 | ING-IND/27_ING-IND/25 | 0.08 | MED/06_MED/04 | 0.20 | MED/36 |
| 0.11 | ING-INF/07_ING-INF/01 | 0.03 | MED/06_MED/08 | 0.09 | MED/39 |
| 0.21 | MED/01_MED/09 | 0.18 | MED/13_MED/09 | 0.36 | VET/09 |
| 0.21 | MED/04_MED/09 | 0.41 | MED/15_MED/09 | | |
| 0.23 | MED/05_MED/04 | 0.27 | MED/18_MED/09 | | |
| 0.28 | MED/05_MED/13 | 0.10 | MED/20_MED/38 | | |
| 0.10 | MED/06_MED/04 | -0.62 | MED/22_MED/18 | | |
| 0.03 | MED/06_MED/08 | -0.41 | MED/22_MED/36 | | |
| -0.03 | MED/06_MED/18 | 0.19 | MED/23_MED/11 | | |
| -0.07 | MED/08_MED/18 | 0.31 | MED/27_MED/26 | | |
| 0.17 | MED/13_MED/09 | 0.08 | MED/29_MED/28 | | |
| 0.43 | MED/15_MED/09 | 0.31 | MED/37_MED/26 | | |
| 0.14 | MED/18_MED/09 | 0.25 | MED/39_MED/26 | | |
| -0.56 | MED/22_MED/18 | 0.41 | MED/46_MED/09 | | |
| 0.10 | MED/23_MED/11 | 0.39 | VET/09_VET/03 | | |
| 0.31 | MED/27_MED/26 | 0.34 | VET/09_VET/08 | | |
| 0.16 | MED/29_MED/28 | | | | |
| -0.12 | MED/36_MED/18 | | | | |
| 0.58 | MED/37_MED/26 | | | | |
| -0.30 | MED/37_MED/27 | | | | |
| 0.27 | MED/39_MED/26 | | | | |
| -0.15 | MED/46_BIO/10 | | | | |
| 0.30 | MED/46_MED/09 | | | | |
| -0.48 | MED/49_BIO/12 | | | | |
| -0.22 | MED/50_MED/28 | | | | |